# Ultrafast Relaxation Dynamics of Photoexcited Dirac Fermion in The Three Dimensional Dirac Semimetal Cadmium Arsenide

Wei Lu,[1] Shaofeng Ge,[1] Xuefeng Liu,[1] Hong Lu,[1] Caizhen Li,[2] Jiawei Lai,[1] Chuan Zhao,[1] Zhimin Liao,[2,3] Shuang Jia,[1,3] and Dong Sun[1,3,*]

[1] *International Center for Quantum Materials, School of Physics, Peking University, Beijing 100871, P. R. China*
[2] *State Key Laboratory for Mesoscopic Physics, Department of Physics, Peking University, Beijing 100871, P.R. China*
[3] *Collaborative Innovation Center of Quantum Matter, Beijing 100871, P. R. China*

Three dimensional (3D) Dirac semimetals which can be seen as 3D analogues of graphene have attracted enormous interests in research recently. In order to apply these ultrahigh-mobility materials in future electronic/optoelectronic devices, it is crucial to understand the relaxation dynamics of photoexcited carriers and their coupling with lattice. In this work, we report ultrafast transient reflection measurements of the photoexcited carrier dynamics in cadmium arsenide ($Cd_3As_2$), which is one of the most stable Dirac semimetals that have been confirmed experimentally. By using low energy probe photon of 0.3 eV, we probed the dynamics of the photoexcited carriers that are Dirac-Fermi-like approaching the Dirac point. We systematically studied the transient reflection on bulk and nanoplate samples that have different doping intensities by tuning the probe wavelength, pump power and lattice temperature, and find that the dynamical evolution of carrier distributions can be retrieved qualitatively by using a two-temperature model. This result is very similar to that of graphene, but the carrier cooling through the optical phonon couplings is slower and lasts over larger electron temperature range because the optical phonon energies in $Cd_3As_2$ are much lower than those in graphene.

---

[*] E-mail: sundong@pku.edu.cn

# I. INTRODUCTION

Massless Dirac Fermions in Dirac semimetals with linear band dispersion mimic the high-energy relativistic particles in their low-energy states [1-9]. The high carrier velocity and the backscattering suppression [5,6,8,10-26] make the mobilities of Dirac semimetals extremely high, which promises revolutionary applications in electronics and optoelectronics with supreme performance. Early realizations of Dirac Fermion are limited in two dimensional (2D) systems: a decade ago with the isolation of 2D graphene and then with the surface of 3D topological insulators [1,2,5,6,10,13,15-17,22]. Recently, prediction and verification of three dimensional (3D) Dirac semimetallic states in pure materials have prompted intense research interest on 3D Dirac semimetals [7,21,24,25,27-30]. Compared to that of graphene and topological insulators, the 3D Dirac semimetallic states are more robust against environmental defects or excess conducting bulk electrons of a 2D surface [21,25,28,31]. Unlike the unstable $BiO_2$ [7] and air-sensitive $A_3Bi$ (A=Na,K,Rb) [21,29,32-34], $Cd_3As_2$ is a stable compound with ultrahigh mobility up to $9 \times 10^6$ cm$^2$ V$^{-1}$ S$^{-1}$, which indicates that it has high-velocity 3D Dirac semimetallic states [23-26,28,35,36]. Its mobility is higher than suspended graphene and among the highest of any bulk semiconductors. Besides the ultrahigh mobility, transport measurements of bulk $Cd_3As_2$ also show giant magnetoresistance [26,37-40], non-trivial quantum oscillations and Landau level splitting under magnetic field. All the results confirm the 3D Dirac semimetal phase in $Cd_3As_2$ [23-25,35-37,41,42]. Furthermore, Dirac semimetals also serve as a starting point to realize Weyl semimetals when the time-reversal or spatial-inversion symmetries are broken [20,28,38,43].

Steady state transport and angle-resolved photoemission spectroscopy (ARPES) measurements mainly confirm the linear band dispersion of the Dirac Fermion and study the behavior of electron near the Fermi level. On the other hand, the transport behaviors in high speed devices, as promised by ultrahigh mobility of Dirac semimetals, are determined by the dynamic conductivity of excited carriers due to the presence of high/dynamic fields in these devices. Therefore, it is crucial to understand electron-electron (e-e) scattering, the cooling of the electrons due to the coupling to lattice phonons and the response of quasiparticle plasma to dynamical fields, if we want to apply the Dirac semimetals in high speed devices. Although the photoexcited Dirac Fermion dynamics are well studied in 2D systems especially that of graphene [44-55], very few experiments have been performed so far on photoexcited Dirac Fermion dynamics in 3D Dirac semimetals. Time resolved ARPES were studies on $ZrTe_5$ [56] and $SrMnBi_2$ [57] respectively. A transient-grating measurement was performed on $Cd_3As_2$ with photon energy of 1.5 eV which is far above the Dirac point and relates to the transitions of multiple bands [58]. In this work, we performed ultrafast transient reflection measurements of photoexcited carrier dynamics close to the Dirac point on both bulk and nanoplate forms of $Cd_3As_2$ and retrieved the dynamical evolution of carrier distributions qualitatively by using a two-temperature model (TTM). The photoexcited carrier dynamics of $Cd_3As_2$ is found to be very similar to that of graphene: after the photoexcitation with 1.5-eV pump photons, photoexcited carriers experience fast thermalization through carrier-carrier scattering and initial relaxation to low energy state by emitting high energy optical phonons with a several ps time constant. The cooling thereafter is dominated by relatively slow, low energy optical phonon and acoustic phonon coupling processes. However, the optical phonon couplings are slower and last over larger electron temperature range, because the optical phonon energies in $Cd_3As_2$ are much lower compared to those in graphene.

## II. METHODS

The bulk $Cd_3As_2$ single crystals were grown out of a Cd-rich melt with the ratio of Cd:As = 85:15 by flux techniques following the same recipe that is described in previous works [42, 59]. $Cd_3As_2$ nanoplates were synthesized from $Cd_3As_2$ powders (Alfa Aesar 99.99% purity) by chemical vapor deposition with a silicon substrate placed at downstream to collect the products [38, 60-62]. Before the growth, the system was flushed several times with Argon gas to thoroughly remove the oxygen in the tube. Then the system was elevated to 650 °C in 20 min under an Ar pressure of 0.8 atm, and held at this condition for 20 min for growth with 20 sccm Ar flow.

The crystalline properties of bulk $Cd_3As_2$ were determined by X-ray diffraction (XRD) measurement (Rigaku MiniFlex 600 diffractometer, with a Cu-Kα1 radiation source). The thickness of $Cd_3As_2$ nanoplate was measured by SPI3800N atomic force microscope (AFM) system with contact mode. Transmission electron microscopy (TEM) and selected area electron diffraction (SAED) characterizations were performed by a FEI Tecnai F20 TEM equipment at 200 keV. For the hall transport measurement, individual nanoplates were transferred onto a silicon substrate with an oxide layer of 285 nm to fabricate devices, and then the measurement was performed in an Oxford cryostat system with magnetic field from 0 T to 14 T and temperature from 10 K to 300 K.

For the transient reflection measurements, a 100 fs, 250 kHz amplified Ti: sapphire laser at 800 nm pumps an infrared optical parametric amplifier (OPA) with signal wavelength tunable from 1.2 to 1.6 μm and idler wavelength tunable from 1.6-2.4 μm. The remnant

800 nm of OPA was compressed back to less than 100 fs and used as the pump. The OPA signal and idler were tuned to 1.34 μm and 2.0 μm with about 150-fs pulse width to pump a difference frequency generator (DFG) to output 4 μm with 220 fs as probe unless in a probe wavelength dependent measurement. The pump and the probe beam are co-linearly polarized. For measurement of the bulk $Cd_3As_2$ crystal, the pump and probe pulses were focused through a 150-mm and 100-mm $CaF_2$ lens respectively onto the sample which was placed in a cryostat for low temperature measurement. The probe spot size is estimated to be about 80 μm while the pump spot size is about 100 μm. The penetration depths at 800-nm pump and 4-μm probe are 1 μm and 2 μm respectively as estimated from absorption measurement results in the literature [63]. The pump power is varied from 1 mW to 10 mW which converts to a photon flux of $0.2\text{-}2\times10^{15}$ photons/cm$^2$. In the detection end, a monochromator is used to select the reflected probe and it is detected by a liquid-nitrogen-cooled InSb photo-detector and the signal is picked up by a lock-in amplifier referenced to 5.7-kHz mechanically chopped pump. The setup of transient reflection measurement of $Cd_3As_2$ nanoplates is similar except a 40× reflective objective lens is used to focus the co-propagating pump probe spots onto the sample. The probe spot size is estimated to be about 6 μm while the pump spot size is about 10 μm. The reflected probe beam is detected by an InGaAs photodiode. Typical pump power used is 0.5 mW unless specified, which converts to a photon flux of $1\times10^{16}$ photons/cm$^2$. The fittings of the transient reflection signals are carried out using Origin Pro 8 software with either mono-exponential or bi-exponential functions as shown below:

$$\Delta R/R = A_0 + A_1 \times \exp(-t/\tau) \text{ and } \Delta R/R = A_0 + A_1 \times \exp(-t/\tau_1) + A_2 \times \exp(-t/\tau_2),$$

where $R$ represents the reflection, $\Delta R$ is differential reflection, $t$ is the pump-probe delay; amplitudes $A_i$ and decay time constants $\tau$ or $\tau_{1(2)}$ are the fitting parameters. The standard error of $\tau$ and $\tau_{1(2)}$ calculated by Origin Pro 8 software is used for the error bar of fitting.

## III. RESULTS AND DISCUSSION

### A. Sample Characteristics

The two categories of $Cd_3As_2$ samples are very different in morphology and doping intensity. Bulk $Cd_3As_2$ crystals are in a needle-like form, as shown in Fig. S1a in the Supplemental Materials [65], the long axis lies along the $[1\bar{1}0]$ direction and the width direction is along $[44\bar{1}]$. The largest facet of the crystal is the (112) plane which is the surface we performed the transient reflection measurements. XRD measurement (Fig. S1b in the Supplemental Materials [65]) confirmed the single-crystalline property of bulk $Cd_3As_2$ [41]. The refined lattice parameters are $a = b = 12.6207(5)$ Å and $c = 25.3756(14)$ Å, which match the literature reports within 0.2% uncertainty [41,59]. The doping is n-type with intensity of about $5.86 \times 10^{18}$ cm$^{-3}$ [42], which converts to a Fermi level of 200 meV above the Dirac point. For the $Cd_3As_2$ nanoplates, the thickness ranges from 200 to 700 nm determined by the AFM measurement and the lateral dimensions of the nanoplates range from several micrometers to tens of micrometers. From the high-resolution TEM measurement (Fig. S1c and S1d in the Supplemental Materials [65]), the inter-planar spacing of the nanoplate is about 0.23 nm, indicating the $(1\bar{1}0)$ edge direction of the nanoplates. The SAED pattern of $Cd_3As_2$ nanoplate (Fig. S1e in the Supplemental Materials [65]) shows a clear set of hexagonal patterns, and the principle bright nodes can be identified as [221], [440] and [40$\bar{8}$] planes, respectively. According to the crystallography calculation, the surface of the nanoplates is (112), which is the plane we performed the transient reflection measurement. The hall measurement at room temperature shows that the Fermi level is about 38 meV above the Dirac point (Fig. S1f in the Supplemental Materials [65]).

## B. Ultrafast Transient Spectroscopy Scheme

Ultrafast transient spectroscopy is one of the standard experimental tools to characterize photoexcited carrier dynamics and high field transport behaviors of condensed matters. As illustrated in the schematic diagram (Fig. 1a and 1b), a 1.5-eV pump pulse excites carriers from the valence to the conduction band across the Dirac point of $Cd_3As_2$, as marked by the transition in the band diagram (Fig. 1c). Here we note that the pump transition labeled in Fig. 1c is one of the many possible transition pathways of 1.5 eV photons since many bands can get involved with such large photon energy. However, after rapid e-e scatterings and subsequently electron-phonon (e-p) scatterings, the pump excited carriers will thermalize and cool down to the lowest conduction and valence bands. During these processes, the energy distributions of the carriers evolve dynamically, which will modify the reflection of the sample. If another pulse probes the pump-induced reflection change, termed as $\Delta R$, at variable delays $t$, the $\Delta R(t)$ signal should correlate to the dynamical evolution of pump excited carrier distributions. In this work, both bulk and nanoplate samples were systematically studied with 2-μm and 4-μm probes. However, since multiple bands are relevant to the 2-μm probe photon transitions and the 2-μm probe transitions are quite above the Dirac point, there are many complexities in data interpretation for 2-μm probe. In the main text, we mainly focus on the discussion of 4-μm probe measurements on the bulk samples, because the 4-μm probe photon corresponds to a simple transition that is closed to the Fermi level of the bulk sample and involves only two bands. In a reflection measurement geometry, the measured $\Delta R$ signal depends on the pump induced changes in both the real and imaginary part of complex refractive index $n(\omega_p)$, thus the response at probe frequency $\omega_p$ should integrate over responses of all optical frequencies according to the Kramers-Kronig relations ideally. However, the contribution from certain frequency $\omega$ decays with the detuning from $\omega_p$ following the relationship $1/(\omega-\omega_p)^2$ and the overall response should still be dominated by the transitions at the close neighbor of 4-μm probe. This is especially true after the initial relaxation from highly nonequilibrium states when dynamical evolutions of carrier distributions are mostly around the Fermi level that is close to the probe transitions. Moreover, as described later, the experimentally observed features with 4-μm probe can be easily interpreted with a two-temperature model [66].

### C. Ultrafast Transient Dyamics Features of Photoexcited Carriers

Figure 2a and 2b show the typical transient reflection results of bulk and nanoplate samples with 2-μm and 4-μm probe at room temperature, respectively. We note the transient responses ($\Delta R/R$) of bulk and nanoplate samples are very different in terms of sign and absolute amplitude, but the overall relaxations of transient response are similar with the same probe wavelength. Compared to that of bulk, the $\Delta R/R$ magnitudes of nanoplate are lower at the same probe wavelength although the pump excitation density is even higher in nanoplate. With 4-μm probe, the dynamical evolutions of $\Delta R/R$ are similar in terms of sign for nanoplate and bulk samples at room temperature: negative immediately after the pump excitation and decays to zero thereafter without sign flip. With 2-μm probe, however, the responses are very different for bulk and nanoplate: for bulk, the $\Delta R/R$ is negative immediately after the pump excitation, which indicates a pump-induced decrease of reflection. After a rapid decay, the signal flips to positive and remains to be positive over the rest of spectral range. In contrast, for nanoplate, the Δ*R*/*R* is positive immediately after the pump excitation, and switches to negative within 2.2 ps and stays negative until decaying to zero. Many factors such as pump photon flux, doping level and morphology, may contribute to the different transient responses of nanoplate and bulk. The pump photon energy and doping level dependence will be discussed in detail later. The morphology difference of the nanoplate and the bulk samples may affect the transient responses according to at least two factors: first, the nanoplate has larger surface to volume ratio compared to that of the bulk, while the surface state would potentially affect the transient dynamics significantly; second, the nanoplate is much thinner than the bulk, and has more efficient heat dissipation to the substrate. However, the heat dissipation process has relatively long decay time constant.

In Figure 2c and 2d, the transient dynamics of bulk at 77 K with probe wavelength of 2000 nm, 2115 nm, 3880 nm and 4000 nm are compared. The magnitude of the transient response increases as the probe transition is getting closer to the Fermi level of the bulk (Fig. 2c). From the spectra plot that is normalized with negative peak at time zero (Fig. 2d), we notice the initial decay is faster when it is probed by shorter wavelength.

Figure 3 shows the effect of pump power on the transient dynamic of bulk at 10 K. As the pump power increases from 1 mW to 10 mW, the magnitude of negative peak around time zero clearly increases, then Δ*R*/*R* decays and crosses zero at different delays ($t_c$) as plotted in Fig. 3c. After that, the Δ*R*/*R* reaches a positive maximum and then gradually relaxes to zero within 40 ps. It is noteworthy that the maximum amplitude of positive Δ*R*/*R* almost stays as a constant regardless of pump power as shown in Fig. 3c and the inset of Fig. 3a. A normalized plot according to $\Delta R/R|_{t=0}$ (Fig. 3b) reveals that it takes longer for photoexcited carriers to relax to zero crossing point with higher pump excitation power. Further analysis indicates the transient signal from negative peak to zero crossing can be well fitted with a mono-exponential decay function (Fig. 3b) with the decay time constant increasing with pump power (Fig. 3d). These tendencies persist at 77 K as shown in Fig. S3 in the Supplemental Materials [65].

Figure 4a shows the normalized lattice temperature dependence of the transient reflection spectra of bulk with the same pump power. As the lattice temperature increases above 77 K, the $\Delta R/R$ signals can no longer cross zero as evidently as that of 10 K and 77 K, and almost turn to be pure negative with positive data points within the noise level only. Further analyses with bi-exponential fittings of the transient responses (Fig. 4b) indicate that the slow time constant $\tau_2$ increases with temperature. The fast time constant $\tau_1$ also increases with temperature below 150K, but it decreases slightly above 150 K. Since the two exponential decay time constants from bi-exponential fittings are fairly close to each other as shown in Fig. 4b, semi-log scale plots are given in Fig. S4 of Supplemental Materials [65] for judgement of the reliability of bi-exponential fittings.

### D. Interpretation of Transient Dyamics

Although there is no general consensus on quantitative model of the optically excited non-equilibrated carriers' behavior in the field on similar questions and a quantitative understanding from first principle is still lacking, we notice that a simple two temperature model can qualitatively interpret the transient measurement results of our bulk sample, especially for the 4-μm probe results which only involve two bands. The TTM involves two characteristic temperatures for electron $T_e$ and lattice $T_L$ has been successfully applied to metal film and graphene for the similar issues [47,67]. As shown in Fig. 5, immediately after the 1.5-eV pump photon excitation and subsequent initial thermalization through e-e scattering, the carriers will reach a quasi-equilibrium distribution that can be characterized by an electron temperature ($T_e$) using Fermi-distribution function as shown in Fig. 5b. This leaves a smeared Fermi surface that is characterized by $T_e$ shortly after pump excitations (~ps). The TTM is different from the Rothwarf-Taylor model that is used to describe a gapped semiconductor [68], where the separated quasi Fermi-distributions of electrons and holes have to be considered before the electron-hole recombination in a nanosecond scale [69,70]. Indeed, there is some evidences showing that during the first picosecond, electrons and holes each maintain a distinct transient chemical potential in graphene [71-74]. The TTM is valid in $Cd_3As_2$ because the electron scattering time is usually much faster than the pulse width (~100 fs) in solids and the photoexcited holes can be instantaneously filled with electrons through efficient e-e scattering due to the gapless structure which is similar to that in graphene and other semimetallic materials [57,72-74].

After the rapid thermalization among electrons, the electron temperature cools down through the e-p scattering processes (Fig. 5c), which include a fast cooling process by emitting optical phonons with relatively large energy and a slow cooling process through scattering with low energy acoustic phonons. The energies of optical phonon branches of $Cd_3As_2$ are around 25 meV (~288K) [75,76], much lower than the optical G phonon of 195 meV in graphene [77]. For this reason, the optical phonon cooling is slower in $Cd_3As_2$ (~2-5 ps) compared to that in graphene (<1 ps). On the other hand, because the optical phonon coupling survives until much lower electron temperature, the electron temperature of $Cd_3As_2$ exhibits faster cooling within the low electron temperature range due to the existence of optical phonon coupling [75]. The theoretically predicted decay time constant through acoustic phonon coupling is on the order of microsecond around the Dirac point and decreases to nanosecond for heavily doped samples. In the presence of short-range disorder, the cooling power is enhanced by a factor of 250, which makes the decay time constant falling into few picosecond region [78]. However, considering the good crystal quality revealed by high-resolution TEM of the nanoplate (Fig. S1d of Supplemental Materials [65]), and the extremely high mobility measured on the bulk samples from the same growth batch [42], the decay time constant in picosecond timescale and the slower decay time constant in a bi-exponential fitting are unlikely merely due to the acoustic phonon coupling.

Compared to the carrier distributions before pump excitation (Fig. 5a), the change after pump excitation and subsequent thermalization process is mainly around the Fermi surface as illustrated in Fig. 5d. There are more carrier-occupy states above the Fermi level and less carrier-occupy states below the Fermi level. This carrier distribution change mainly accounts for the transient reflection signal probed by the corresponding probe photon transitions. From this diagram, if the probe transition is closer to the Fermi level of unexcited distributions, larger $|\Delta R/R|$ signal should be observed. This immediately explains why the magnitude of $\Delta R/R$ is larger with 4-μm probe on the bulk sample comparing to that with larger probe photon energies (Fig. 2c) and that on the nanoplate sample with lower Fermi level (Fig. 2a and 2b). Although many factors such as pump photon flux and morphology may contribute to the different transient responses of nanoplate and bulk observed in Fig. 2a and 2b, we emphasize that the doping intensity difference between bulk and nanoplate probably accounts for the experimentally observed difference of transient responses, because the probe transitions are much closer to the Fermi level of heavily doped bulk (200 meV) comparing to weakly doped nanoplate (38 meV).

Because both the absorption and the real part of refractive index contributes to the reflection signal [58], we cannot deduce the differential reflection directly from the pump induced carrier distribution change shown in Fig. 5c based on a simple Pauli blocking picture. However, a qualitative reflection curve as function of instantaneous $T_e$ can be plotted in Fig. 5e with the temperature dependent measurement results (Fig. 4). Four main features are noteworthy in Fig. 5e. First, as $\Delta R$ is negative around time zero and gradually decays, $R(T_e^i) < R(T_L)$; second, $R(T_e)$ decreases with $T_e$ when $T_e > 77$ K according to the $\Delta R/R$ curve at 77 K; third, the $\Delta R/R$ signal crosses zero at $T_e^C$ where $R(T_e^C) = R(T_L)$ from the $\Delta R/R$ curve of 10 K; fourth, $\Delta R/R$ reaches positive maximum at $T_e^M$ when $T_e^M \leq 77$ K, while the $\Delta R/R$ curve at temperature above 77 K does not evidently cross zero anymore. The existence of $T_e^M$ directly explains the almost flat peak amplitudes of positive $\Delta R/R$ at different pump power shown in Fig. 3c: the same positive maximums are reached when the electron temperature cools down to $T_e^M$ regardless of the pump excitation density and initial electron temperatures.

As shown in Fig. 5e, $T_e^M$ and $T_e^C$ can be directly extracted from the transient reflection curves, which helps to fix the characteristic electron temperature range for the relaxation dynamics fittings of $T_e$. If it takes 5-7 ps to relax from $T_e^C$ to $T_e^M$, and $T_e^M \leq 77$ K, we speculate $T_e^C > 300$ K, thus the relaxation from $T_e^i$ to $T_e^C$ should be dominated by optical phonon couplings. The mono-exponential fitting in Fig. 3b indicates the decay time constant increases with pump power when the electron temperature cools from $T_e^i$ to $T_e^C$.

As $T_e^i$ is determined by the pump power in pump power dependent measurements, longer relaxation time from higher $T_e^i$ with higher pump excitation to the fixed $T_e^c$ is expected. That explains the power dependence of $t_c$ observed in Fig. 3c. The increase of the decay time constant with increasing pump power (Fig. 3d) is mainly due to the elevated phonon temperature because both the e-p coupling strength and lattice heat capacity are temperature dependent. Although the heat capacity of lattice is orders larger than that of electrons, the phonon temperature still increases slightly as the electrons dissipate heat to the phonons through e-p couplings. We note the slightly elevated phonon temperature difference at different pump power also accounts for the very weak dependence of ∆R/R ($T_e^M$) on pump power shown in Fig. 3c.

In temperature dependent measurements, $T_e^i$ is mainly determined by the pump excitation with only weak dependence on the initial lattice temperatures when the pump photon flux is large (Fig. S5 and S6 in the Supplemental Materials [65]). As the lattice temperature increases, the relaxation gets slower (Fig. 4b), which has to correlate to the temperature dependence of electron and phonon heat capacities and the e-p coupling strength. To fit the results at 10 K, a mono-exponential decay can fit the range from negative peak to $T_e^c$ with a decay time constant 3.17 ps, similar to that is used in the power dependent measurement. However, because lower energy optical phonon and acoustic phonon coupling certainly play a role during the relaxation [78], the mono-exponential fitting is not sufficient for the cooling after $T_e^c$, while a bi-exponential fitting is used to fit the full transient reflection dynamics of temperature dependent measurement results as shown in Fig. 4. Furthermore, we find that although the bi-exponential function can fit the 77 K, 150 K and 300 K data, the 10 K data in both Fig. 3 and 4 require tri-exponential rather than bi-exponential to be fit well as shown in Fig. S2 of Supplemental Materials [65]. The two exponential components from the bi-exponential fitting do not necessarily correspond to the optical phonon coupling and acoustic phonon decay processes respectively. The acoustic phonon cooling process may not be resolved in our measurement due to the limited signal to noise ratio, while the decay time constant should be far longer if the theoretical prediction is correct [78]. Thus the two decay constants may be dominated by the optical phonon coupling with different branches, considering the rich optical phonon modes available in $Cd_3As_2$ [75]. Alternatively, it is simply a magnification of lattice heat capacity variation over different temperature ranges.

## VI. CONCLUSIONS

In conclusion, the photoexcited carrier dynamics of $Cd_3As_2$ are investigated using ultrafast transient reflection measurements with probe photon transitions close to the Dirac point and Fermi level. We found that the carrier dynamics of $Cd_3As_2$ can be qualitatively interpreted with a two-temperature model similar to that of graphene, but the optical phonon couplings are slower and last over larger electron temperature range. The fast transient time of the photoexcited carriers observed in this work promises $Cd_3As_2$ as an excellent candidate for ultrafast optoelectronics and photonics applications such as ultrafast photodetectors [79], optical switches [80] and saturate absorbers, especially for those working in the challenging middle/far IR and THz wavelength range.

## ACKNOWLEDGEMENTS

The first two authors W. Lu and S. Ge contributed equally to this work. This project has been supported by the National Key Research and Development Program of China (Grant No. 2016YFA0300802), National Basic Research Program of China (973 Grant No. 2014CB920900), the National Natural Science Foundation of China (NSFC Grant No.

11674013 and No. 11274015), the Recruitment Program of Global Experts and the Beijing Natural Science Foundation (Grant No. 4142024).

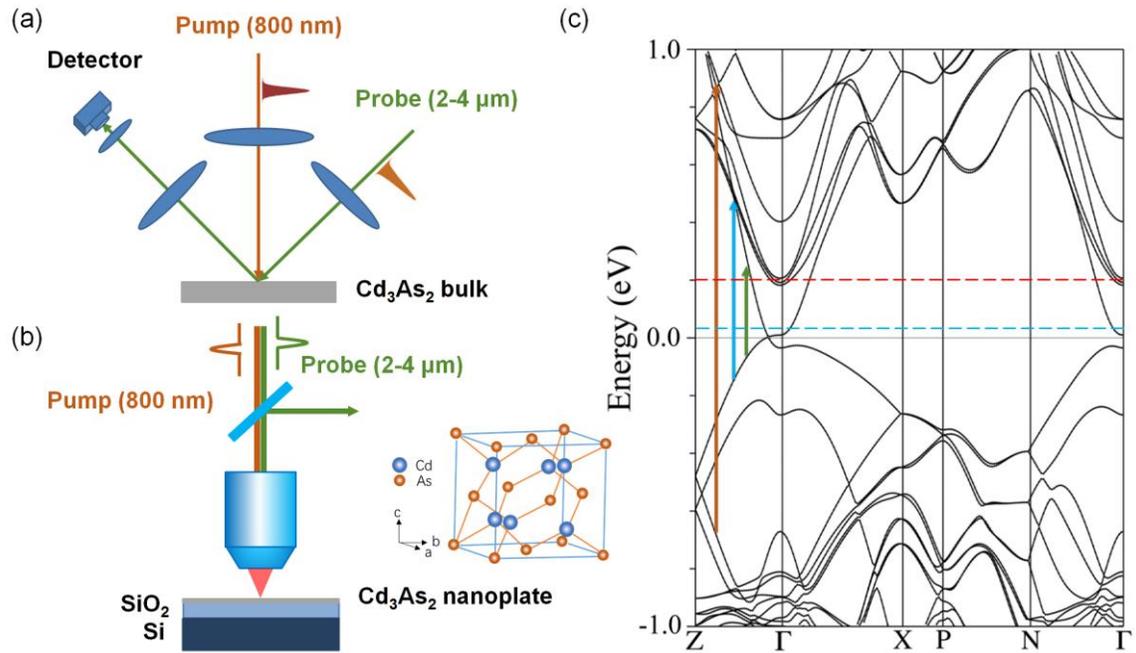

FIG. 1. Schematic diagrams of transient reflection experiment on (a) a bulk crystal and (b) a nanoplate. Both samples are pumped by 800 nm and probed by 2 μm and 4 μm unless in a probe wavelength dependent measurements. Reproduced with permission [64]. Copyright 2015, American Chemical Society. (c) Band diagram of $Cd_3As_2$ and pump/probe photon transition configuration, the Fermi levels of the bulk and nanoplate $Cd_3As_2$ are denoted as red and blue dash lines, and the 800 nm, 2 μm and 4 μm photon relevant transitions are denoted as red, blue and green arrows, respectively. Reproduced with permission [59]. Copyright 2015, American Chemical Society. Inset shows the crystal structure of the $Cd_3As_2$.

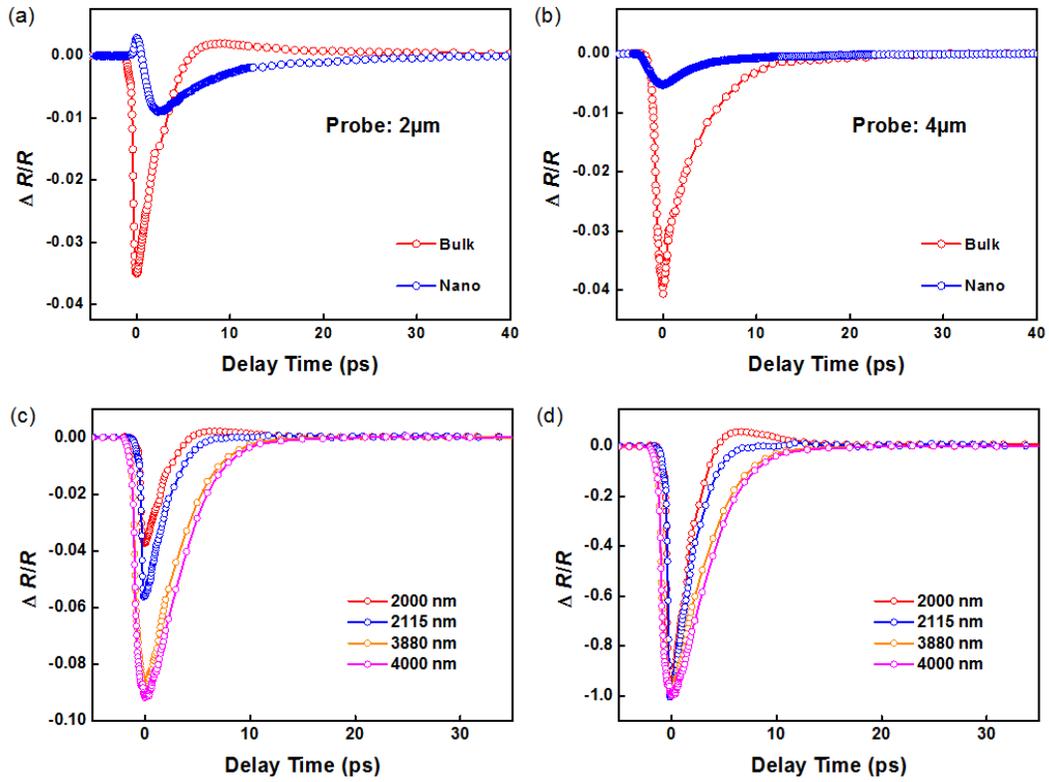

FIG. 2. The typical transient reflection spectra of $Cd_3As_2$ bulk sample and nanoplate at room temperature for (a) 2 μm probe and (b) 4 μm probe respectively. (c) Probe wavelength dependence of transient reflection spectra for 2000 nm, 2115 nm, 3880 nm and 4000 nm on bulk $Cd_3As_2$ at 77 K. The pump power is 5 mW. (d) Normalized plot of Fig. 2(c) according to the negative peak signals at time zero ($\Delta R/R|_{t=0}$).

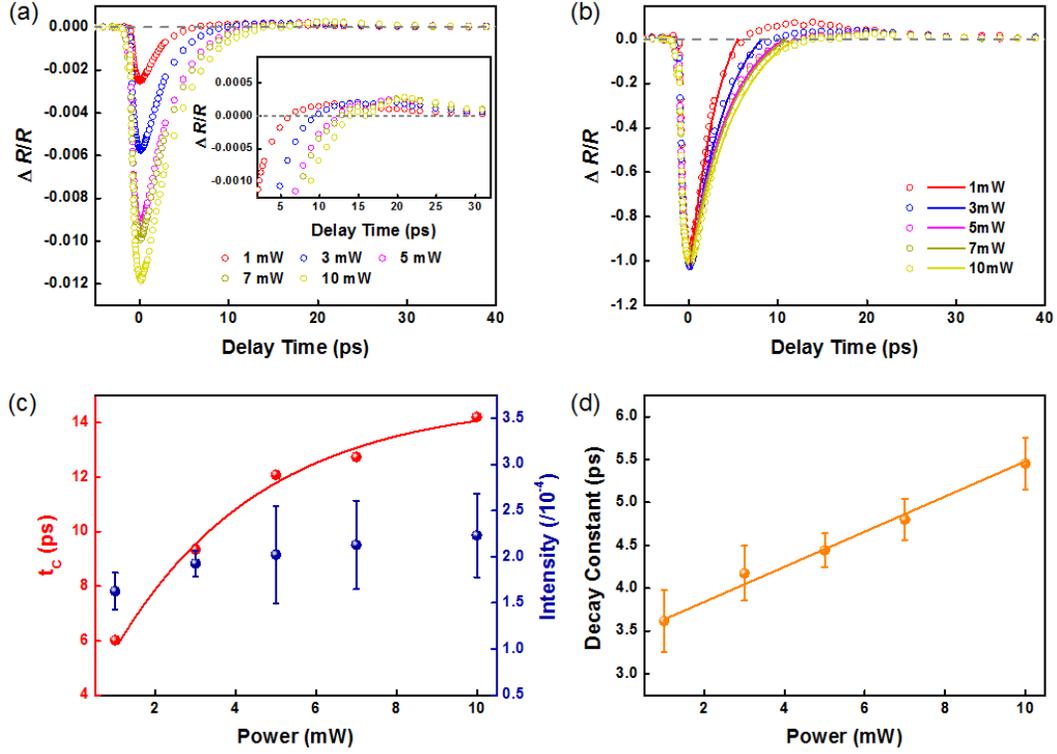

FIG. 3. (a) Pump power dependent transient reflection spectra for 4 μm probe on bulk $Cd_3As_2$ at 10 K. The pump power is varied from 1 mW to 10 mW. The inset magnifies the positive $\Delta R/R$ signals. (b) Normalized plot of Fig. 3(a) according to the negative peak signals at time zero. The spectra are fitted by exponential decay function: $\Delta R/R = A_0 + A_1 \times \exp(-t/\tau)$ from time zero to the delay times ($t_c$) that cross zero with $\tau$ as the decay constant. (c) Pump power dependence of $t_c$ (red) and maximum positive signals (dark blue). (d) Pump power dependence of the decay constant $\tau$. The lines in Fig. 3(c) and 3(d) are guide to the eye.

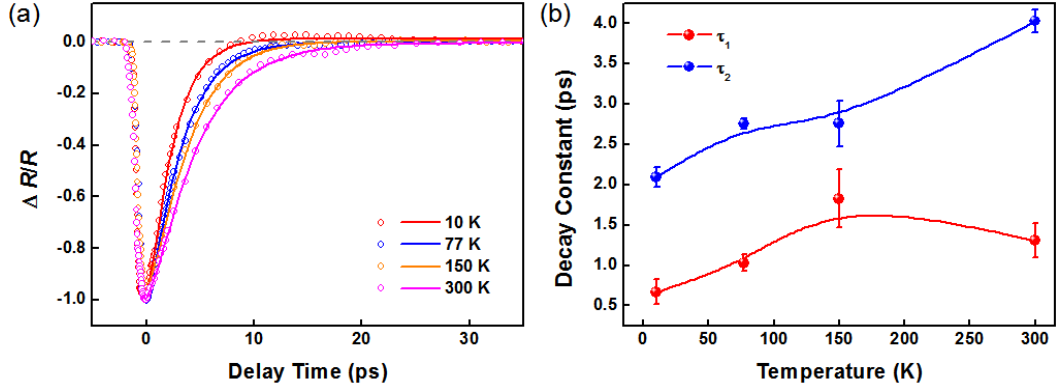

FIG. 4. (a) Normalized temperature dependent transient reflection spectra according to the negative peak signals at time zero for 4 μm probe on bulk $Cd_3As_2$ sample at 10 K, 78 K, 150 K and 300 K, respectively. The pump power is 5 mW. The decay of $\Delta R/R$ from the negative peaks are fitted by a bi-exponential decay function: $\Delta R/R = A_0 + A_1 \times \exp(-t/\tau_1) + A_2 \times \exp(-t/\tau_2)$ with a fast time constant $\tau_1$ and a slow time constant $\tau_2$. (b) The temperature dependence of $\tau_1$ and $\tau_2$ extracted from the fitting of Fig. 4(a), the lines are guide to the eye.

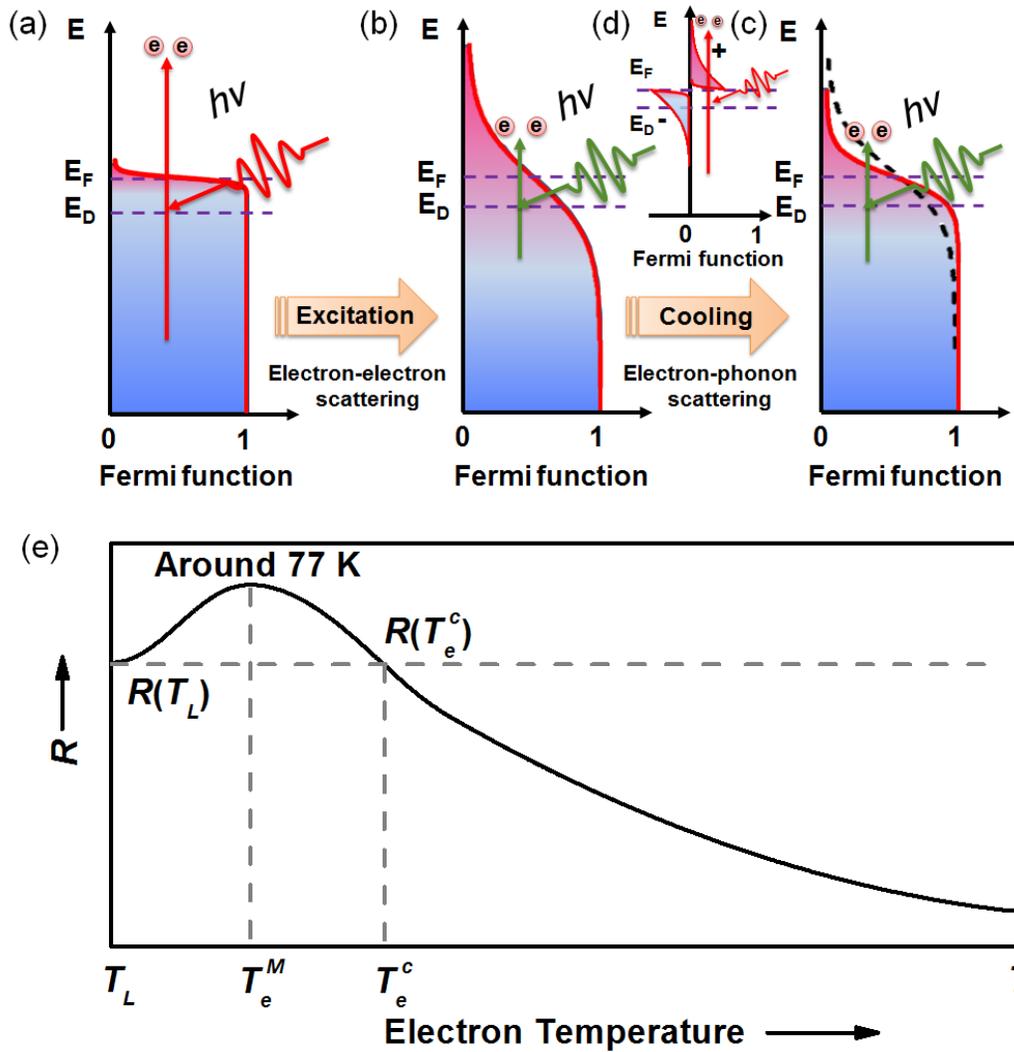

FIG. 5. Schematic diagrams of the carrier distribution around the Fermi level. (a) Before the pump excitation, (b) after pump excitation and initial thermalization through e-e scattering, and (c) after cooling with e-p scattering. (d) The carrier distribution difference of Fig. 5(a) and 5(b). (e) Transient reflectivity of bulk at 4 μm as function of $T_e$ deduced from the temperature dependent measurement (Fig. 4).